\documentclass[twocolumn,secnumarabic,amssymb, nobibnotes, aps, prb, superscriptaddress]{revtex4-2}
\usepackage{lipsum}
\usepackage{titlesec}
\titlespacing\section{0pt}{12pt plus 4pt minus 4pt}{4pt plus 20pt minus 2pt}
\usepackage{xcolor}
\usepackage{braket}
\usepackage{amsmath}
\usepackage{comment}
\usepackage{physics}
\usepackage{afterpage}
\usepackage{placeins}
\usepackage{graphicx}
\usepackage{float}
\usepackage{cancel}
\usepackage{booktabs}
\usepackage{ulem}
\usepackage{multirow}
\usepackage{array}
\usepackage{setspace}
\graphicspath{{Figs/}}
\usepackage{siunitx}
\usepackage{hhline}
\usepackage{xfrac}
\usepackage{float,graphicx}
\usepackage{mathtools}
\usepackage{listings}
\usepackage{amssymb}
\usepackage{soul}
\usepackage{hyperref}
\usepackage{titlesec}
\usepackage{amsfonts}
\usepackage[version=4]{mhchem}

\usepackage{epstopdf}
\DeclareUnicodeCharacter{0308}{HERE!HERE!}
\catcode`@11
\def\seceqaa{\@addtoreset{equation}{section}
\def\theequation{A\arabic{equation}}}
\def\seceqbb{\@addtoreset{equation}{section}
\def\theequation{B\arabic{equation}}}
\def\seceqcc{\@addtoreset{equation}{section}
\def\theequation{C\arabic{equation}}}
\def\seceqdd{\@addtoreset{equation}{section}
\def\theequation{D\arabic{equation}}}
\def\seceqee{\@addtoreset{equation}{section}
\def\theequation{E\arabic{equation}}}
\def\seceqff{\@addtoreset{equation}{section}
\def\theequation{F\arabic{equation}}}
\def\seceqgg{\@addtoreset{equation}{section}
\def\theequation{G\arabic{equation}}}
\def\seceqhh{\@addtoreset{equation}{section}
\def\theequation{H\arabic{equation}}}
\catcode`@11

\begin{document}

\title{Berry curvature-induced intrinsic spin Hall effect in light-element-based CrN system for magnetization switching} 

\author{Gaurav K. Shukla\textsuperscript{*}}
\affiliation{National Institute for Materials Science, Tsukuba Japan}
\author{Prabhat Kumar}
\affiliation{National Institute for Materials Science, Tsukuba Japan}
\author{Shinji Isogami\textsuperscript{\dag}}
\affiliation{National Institute for Materials Science, Tsukuba Japan}


\begin{abstract}
 The current-induced spin-orbit torque-based devices for magnetization switching are commonly relied on the $4d$ and $5d$ heavy metals owing to their strong spin-orbit coupling (SOC) to produce large spin current via spin Hall effect (SHE). Here we present the sizable SHE in CrN, a light-element-based system and demonstrate the current-induced magnetization switching in the adjacent ferromagnetic layer [Co(0.35\,nm)/Pt(0.3\,nm)]$_3$, which exhibits perpendicular magnetic anisotropy. We found the switching current density of 2.6 MA/cm$^2$. The first principles calculation gives the spin Hall conductivity (SHC) $\sim$ $120\, (\frac{\hbar}{e})$\,S/cm due to intrinsic Berry curvature arising from SOC induced band splitting near Fermi-energy. The theoretically calculated intrinsic SHC is close to the experimental SHC extracted from second harmonic Hall measurement. We estimated spin Hall angle ($\theta$$_{SH}$) $\sim$ 0.09, demonstrating efficient charge-to-spin conversion in CrN system. 
 \end{abstract}

\maketitle
\section{INTRODUCTION}
Spintronics revolutionized the area of data storage technology harnessing the electron spin in role instead of charge current \cite{bader2010spintronics,dieny2020opportunities}. Spin Hall effect (SHE), where an unpolarized charge current flowing through high spin-orbit coupled material is converted into transverse spin-polarized current, has attracted significant interest in spintronics as an efficient way to generate spin current \cite{sinova2015spin,jungwirth2012spin}. 
SHE has two kinds of origin, one is intrinsic mechanism related to the spin Berry curvature associated with Bloch bands and extrinsic mechanism related to the scattering events \cite{zhang2021different}. 
The spin Hall conductivity (SHC) and spin Hall angle ($\theta_{SH}$) are two factors that determines the performance of SHE-based devices \cite{zhao2022large} (Here, SHE is defined as a comprehensive term that includes both SHC and $\theta_{SH}$ in this study). The SHC is the sum of the spin Berry curvature of the occupied states in the Brillouin zone, while $\theta_{SH}$ is the ratio of SHC to the charge conductivity ($\theta_{SH}= \frac{\sigma_{SH}}{\sigma_{xx}}$), that describes the charge-to-spin conversion efficiency \cite{qiao2018calculation,sinova2015spin,guo2008intrinsic}.

The spin current generated by the SHE can exert torque on an adjacent ferromagnetic layer, which is known as spin-orbit torque (SOT), and the magnetization switching by the SOT is a key mechanism for next-generation magnetoresistive random access memory (MRAM) devices \cite{shao2021roadmap,song2021spin,ramaswamy2018recent,han2021spin}. So far, SOT research has primarily concentrated on heavy-metal (HM)/ferromagnetic (FM) heterostructures \cite{song2021spin,hayashi2021spin,peng2019modulation,akyol2016effect} due to the large SHE of HMs such as W, Ta and Pt, arising from high spin-orbit coupling (SOC) \cite{sinova2015spin,wang2014scaling}. On the other hand, there has been growing interest on light $3d$ metals with low SOC such as Cr and V, as a candidate for replacing the HMs in terms of their environmental compatibility and cost-effectivity\cite{guo2021current,chuang2019cr,gupta2025harnessing}. However, the SHE of light $3d$ metals is lower than that of HMs in general, which is an issue to be solved and a challenge for their use in energy-efficient SOT devices \cite{zheng2025spin,wang2014scaling,du2014systematic}. 

In recent years, Nitrogen (N)-based systems widely garnered attention due to their higher stability, long spin diffusion length, topological band feature and seamless hybridization with the light $3d$ metals, enabling the manipulation of their intrinsic properties
\cite{isogami2023antiperovskite,tripathi2024impact,swatek2022room,nandi2024intrinsic}. In addition, it is reported that N also offers a significant impact on the SHE of antiferromagnets \cite{tripathi2024impact}. Thus, the N incorporation is expected to enhance the SHE of light $3d$ metals even though the SOC is low, which may lead to the device applications.

Although $\theta_{SH}$ of Cr, which is one of the light $3d$ metals, is half size of Pt \cite{wang2014scaling,du2014systematic,chuang2019cr}, current-induced magnetization switching (CIMS) has been demonstrated in the Cr-based SOT device, Cr/CoFeB/MgO/Cr \cite{chuang2019cr}. 
In this study, therefore, we focused on CrN, a light-element-based system, where incorporating N into Cr. The CrN is a stable compound with a rock salt structure (NaCl-type, space group Fm$\bar3$m) and paramagnetic structure at room temperature \cite{azouaoui2020structural,biswas2023magnetic}, whose electrical resistivity is reported to be ranging from 0.75 to 300\,m$\Omega$\,cm \cite{gharavi2018microstructure}. Below the Néel temperature (272\,K-280\,K), transformation to the antiferromagnetic state takes place with an orthorhombic crystal structure \cite{biswas2023magnetic}.

In this work, we experimentally demonstrated CIMS in a Co/Pt ferromagnetic multilayer due to spin current from CrN, using the CrN-based SOT device. We measured the switching current density to be 2.6 MA/cm$^2$ in the device. Theoretical calculation gives the intrinsic SHC arising from Berry curvature-induced band splitting near the Fermi energy. The theoretically calculated intrinsic SHC is close to the experimental SHC extracted from second harmonic Hall measurement. We estimated $\theta_{SH}$ of CrN $\sim$ 0.09, which surpasses the $\theta_{SH}$ of Cr ($\sim$ 0.05) \cite{du2014systematic}, demonstrating its efficient charge-to-spin conversion. 
\section{METHODS}
CrN(5)/[Co(0.35)/Pt(0.3)]$_3$/MgO(3) (here parenthesis denotes nominal thickness in nm)  multilayer is grown on $c$-plane oriented Al$_2$O$_3$ substrate using DC and RF magnetron sputtering technique \cite{kelly2000magnetron}. The CrN was deposited using direct current power source at 650\,$^{\circ}$C. The Co, Pt and MgO layers were deposited using radio-frequency power source at room temperature. The base pressure of the chamber was better than 6$\times$10$^{-6}$ Pa. The thin film was patterned into Hall bar device using photolithography and Argon (Ar) ion etching technique. The gold (Au) was sputtered for electrical contact. X-ray diffraction measurement ($\lambda$ = 1.54\AA) was performed for structural analysis. The magnetization measurement was performed using Magnetic Properties Measurement System (MPMS). The anomalous Hall effect  and the CIMS measurements were done using home-build set-up. For CIMS, initially a pulse current with a width of 10\,ms was applied using  a pulse generator. A 0.2\,mA of sensing current was applied from DC power source and the Hall voltage was read using digital multimeter after each interval of current pulse for 1 sec. The magnetic field was applied by an electromagnet. The second-harmonic Hall voltage was measured with a lock-in amplifier while varying in-plane magnetic field. A sinusoidal wave with an effective amplitude of 2.3 mA (current density $\sim$ 3.3 MA/cm²) and a frequency of 33.123 Hz was generated using a pulse generator. The same device and sample package were utilized for both CIMS and second-harmonic Hall measurements. All the measurements were conducted at room temperature. The electronic structure of CrN was calculated employing density functional theory (DFT) using Quantum Espresso package \cite{giannozzi2009quantum}. We employed the Perdew-Burke-Ernzerhof (PBE) type generalized gradient approximation (GGA) for the exchange-correlation functional \cite{ernzerhof1999assessment}. The kinetic energy cutoff of 80\,Ry was taken for the plane-wave basis. A 8 × 8 × 8 $k$-point mesh was used for the Brillouin zone (BZ) sampling. The onsite coulomb interaction ($U$) of 3.0\,eV for Cr atom is taken in the calculation. The spin transport properties were calculated employing the maximally localized Wannier functions using the Wannier90 code \cite{haber2023maximally}. 
\begin{figure*}[t]
    \centering
    \includegraphics[width=0.9\textwidth]{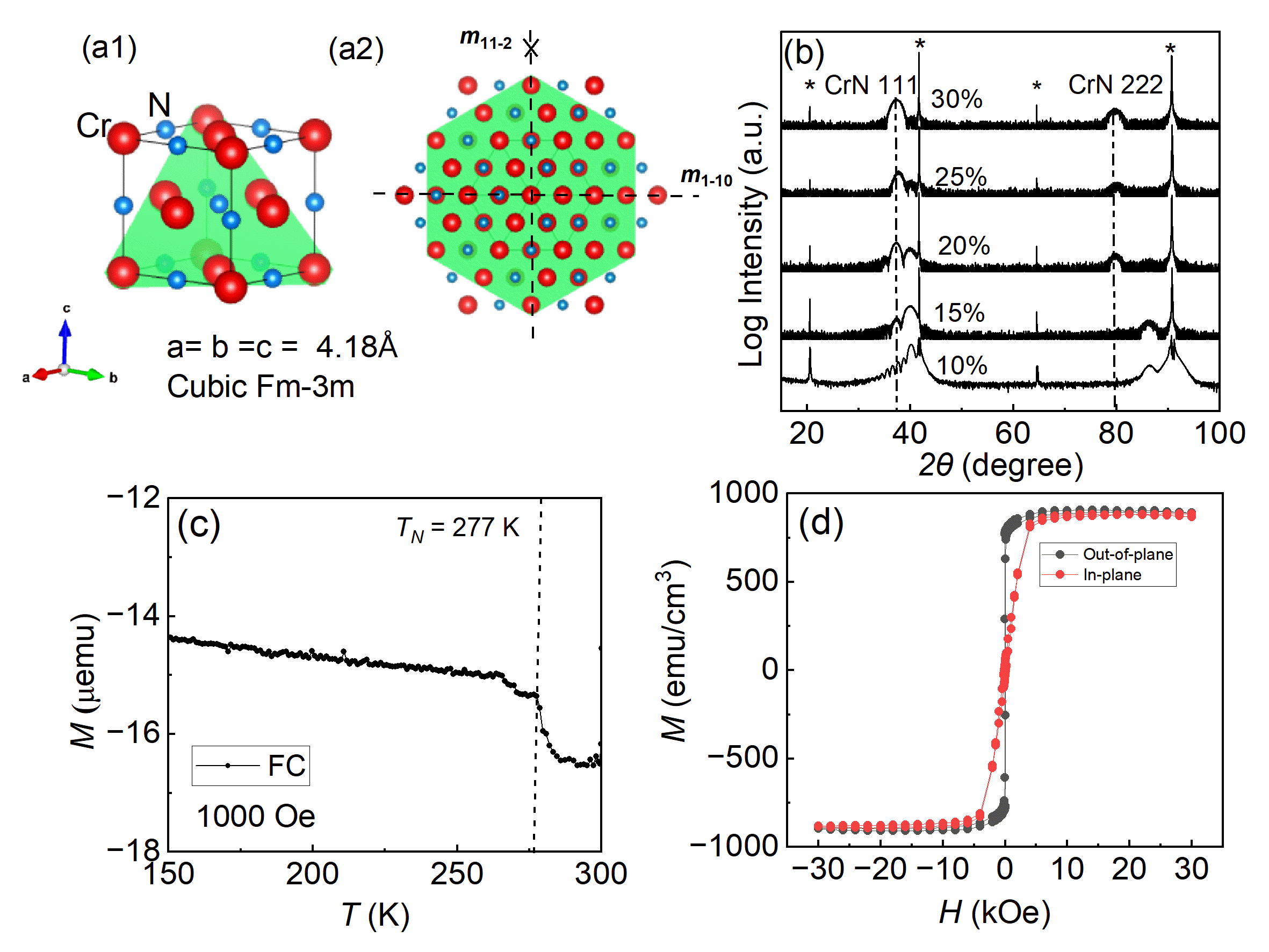}
    \caption{(a1) A face-centered cubic unit cell (space group Fm$\bar3$m) of CrN. Red and light blue spheres represent the Cr and N atoms, respectively. (a2) The (111)-oriented view of CrN unit cell.  The mirror symmetry is preserved along [1-10] direction, while broken along [11-2] direction (dashed lines). The green shaded region in Figs.(a1) and (a2) shows the (111) plane. (b) X-ray diffraction patterns at different N$_2$ flow rate. (*) denotes the substrate peaks. (c) Field cooled (FC) magnetization curve for Al$_2$O$_3$/CrN(5) in the temperature range of 150\,K to 300\,K at 1000 Oe. (d) Out-of-plane (black curve) and in-plane (red curve) magnetization of CrN(5)/[Co (0.35)/Pt(0.3)]$_3$/MgO(3) multilayer device.}
    \label{Fig1}
\end{figure*}
\begin{figure*}[t]
\centering
\includegraphics[width=1\textwidth]{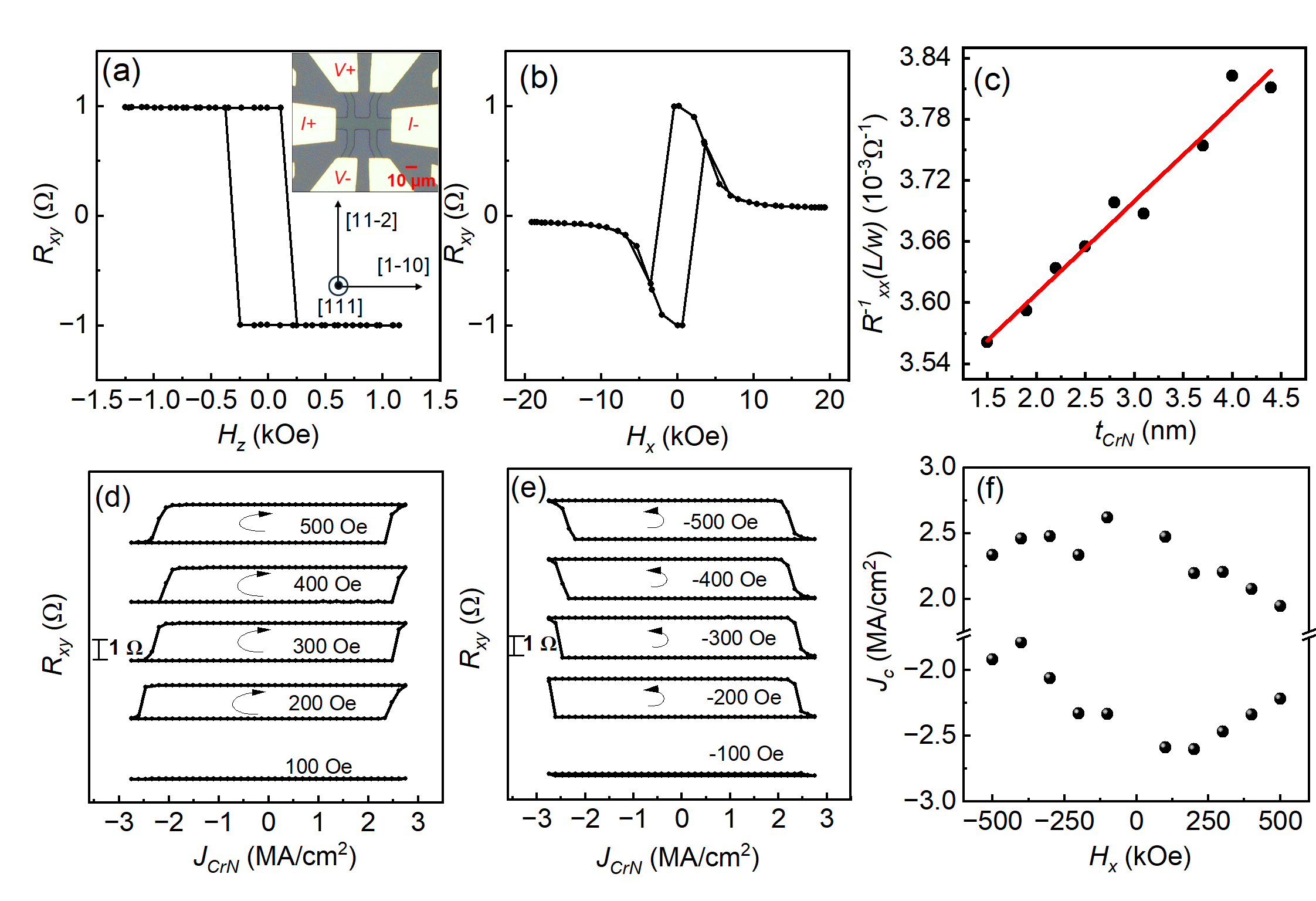}
\caption{(a) Anomalous Hall resistance ($R_{xy}$) as a function of out-of-plane magnetic field ($H$//[111]). Inset shows an optical microscope image of the actual device. (b) Anomalous Hall resistance ($R_{xy}$) as a function of in-plane magnetic field ($H$//[1-10]). (c) The inverse of sheet resistance $L/(R_{xx}w)$ versus CrN layer thickness $t_{CrN}$ plot (black balls). Red line represents the fitted data to extract resistivity of each layer. Current-induced magnetization switching (CIMS) curves as function of current density through CrN layer at (d) positive and (e) negative in-plane external magnetic fields. (f) Variation of switching current density $J_c$ for different in-plane magnetic fields values.}
\label{Fig2}
\end{figure*}
The Kubo formula implemented in Wannier90 code is used to calculate the $k$-resolved Berry curvature, which is given as \cite{guo2008intrinsic,yao2004first}
\begin{equation}
   \Omega_{n,ij}(k) = {\hbar}^2Im\sum_{m\not=n}\frac{-2\mel{nk}{\hat{j^s_{i}}}{mk}\mel{mk}{\hat{\nu_{j}}}{nk}}{(\epsilon_{nk}-\epsilon_{mk})^{2}},
   \label{eq:6}
\end{equation}
where $\hat{j^s_{i}} = \frac{1}{2}\{\sigma_z,v_i \}$ represents the spin-current operator.

The intrinsic SHC is calculated using following formulae\cite{zhou2019intrinsic}
\begin{equation}
   \sigma_{ij}^{k}= -\frac{e^2}{\hbar}\frac{1}{V{N_{k}^3}}\sum_{n}\sum_{k} \Omega_{n,ij}(k) f_{nk},
   \label{eq:7}
\end{equation}
where $V$ is the primitive cell volume and ${N_{k}^3}$ is the number of \textit{k}-points in the Brillouin zone. $f_{nk}$ is the Fermi-Dirac distribution function.
\section{RESULTS AND DISCUSSION}
Figure\,\ref{Fig1}(a1) shows a three-dimensional unit-cell of CrN. Crystallographic studies suggest that CrN belongs to face-centered cubic (FCC) lattice (Fm$\bar3$m space group), with Cr (red spheres) and N (light blue spheres) atoms at (0,\,0,\,0) and (0.5\,,0.5\,,0.5) Wyckoff positions, respectively \cite{kalal2023electronic}.  The (111)-oriented view of CrN is shown in Fig.\ref{Fig1}(a2). In this orientation, CrN retains a mirror symmetry along the [1-10] direction ($m_{1-10}$), while the mirror symmetry along [11-2] ($m_{11-2}$) is broken. The green shaded region in Figs.\ref{Fig1}a(1) and a(2) show the (111) plane. Figure\,\ref{Fig1}(b) shows the room-temperature X-ray diffraction patterns of the grown samples, where different N$_2$ flow rate was introduced for the phase optimization of 5\,nm thick CrN on $c$-plane oriented Al$_2$O$_3$ substrate. At an initial N$_2$ flow rate of 10\% (Ar: 90 sccm,N$_2$: 10 sccm), Cr$_2$N is observed as the dominant phase, with a characteristic peak at $2\theta$\,$\sim$\,40.15 degree.  As the N$_2$ flow rate increases, the Cr$_2$N phase diminishes, and the CrN phase begins to form. At  30\% N$_2$ flow rate, a pure CrN phase is observed. CrN crystallizes in a (111) oriented FCC cubic lattice on Al$_2$O$_3$ substrate, with a lattice parameter $\sim$ 4.18\,\AA, as determined by XRD data.
To determine the transition temperature of CrN, we measured the field cooled (FC) magnetization of Al$_2$O$_3$/CrN(5) from 150\,K to 300\,K at 1000 Oe (Fig.\ref{Fig1}(c)). The antiferromagnetic-to-paramagnetic transition ($T_N$) is observed at 277\,K, which is consistent with the literature. Noteworthy, the negative value of magnetization is due to contribution from Al$_2$O$_3$ substrate \cite{alam2024synthesis,biswas2023magnetic}.  

Following the confirmation of the optimization conditions for CrN\,(111) phase, we synthesized  CrN(5)/[Co(0.35)/Pt(0.3)]$_3$/MgO(3) multilayer device on $c$-plane oriented Al$_2$O$_3$ substrate. Figure\,\ref{Fig1}(d) presents the magnetic hysteresis loop of the device measured at room temperature, where the magnetic field is varied along out-of-plane (black curve) and in-plane (red curve) directions. The out-of-plane magnetization saturates at a much lower field compared to the in-plane magnetization, indicating easy-axis along out-of-plane direction. The unidirectional magnetic anisotropy ($K_u$) was estimated using the in-plane and out-of plane $M-H$ curves using following equation \cite{isogami2024noncoplanar}
\begin{equation}\label{eq:1}
    K_u = K^{eff}_u + 2\pi M_s^2
\end{equation}
\begin{equation}
    K^{eff}_u = \left(\mu_0\int_{0}^{M_s}Hdm\right)_{hard\,axis}  - \left(\mu_0\int_{0}^{M_s}Hdm\right)_{easy\,axis} 
\end{equation}
where the second term of Eq.\ref{eq:1} represents the demagnetization component. Using saturation magnetization ($M_s$) of 890 emu/cm$^3$, $K_u$ was found to be $\sim$ 8.07$\times$\,10$^6$  erg/cm$^3$. This value suggest that device has significant preference of the along out-of-plane direction, which is crucial for SOT driven magnetization switching.  
Now, we investigate the transport properties of the fabricated multilayer device. For the magnetization switching it is crucial that ferromagnetic layer on top of spin-current source should show the perpendicular magnetic anisotropy (PMA). To measure this, we fabricated the Hall bar device with current channel along [1-10] direction (along the mirror symmetry line shown in Fig.\ref{Fig1}(a2)) and voltage channel along [11-2] direction. The length ($L$) and width ($w$) of the current channel are 25\,\textmu m and 10\,\textmu m, respectively. An optical microscope image of the Hall bar device is shown in the inset of Fig.\ref{Fig2}(a). This geometry allows to detect the magnetization state via anomalous Hall effect (AHE). AHE is a phenomenon, where the electric current through the ferromagnetic material generates the transverse voltage due to its magnetization \cite{shukla2021anomalous,nagaosa2010anomalous}. 
The anomalous Hall resistance ($R_{xy}$) of device as a function of out-of-plane external magnetic field ($H$//[111] and $I$//[1-10]) is shown in Fig.\ref{Fig2}(a). A square-shape hysteresis loop as well as sharp transition of $R_{xy}$ shows the easy-axis along [111] direction. The coercivity, which is measure of resistance to change the magnetization state is measured to be 247\,Oe. Figure\,\ref{Fig2}(b) shows the $R_{xy}$ as a function of in-plane magnetic field ($H$ and $I$//[1-10]). The $R_{xy}$ saturates at 15\,kOe suggesting the [1-10] direction is hard axis of magnetization. 
After confirming the PMA in the device, we estimated the resistivity of each conducting layer to measure the actual current passing through CrN layer. For this, we fabricated a multilayer device with CrN thickness varying from 0 to 5\,nm i.e. Al$_2$O$_3$/CrN(0-5)/[Co (0.35)/Pt(0.3)]$_3$/MgO(3) and measured the thickness dependent longitudinal resistance ($R_{xx}$). The $R_{xx}$ was measured in four-point probe geometry.  The inverse of the sheet resistance $i.e.$ $L/(R_{xx}w)$ of the multilayer device as a function of CrN thickness ($t_{CrN}$) is shown in Fig.\ref{Fig2}(c). The linear fitting of this data using equation $ \frac{L}{R_{xx}w} =\frac{t_{CrN}}{\rho_{CrN}} + \frac{t_{Co/Pt}}{\rho_{Co/Pt}}$ (red line) gives $\rho_{CrN}$ and $\rho_{Co/Pt}$ of 1.09\,m$\Omega$\,cm and  57 \textmu$\Omega$\,cm, respectively. The value of the $\rho_{CrN}$ is consistent with the reported in literature \cite{gharavi2018microstructure,biswas2023magnetic}. 

We employ the CIMS measurement in the presence of in-plane external magnetic field ($H$//[1-10]). Figure\ref{Fig2}(d) shows the current-induced magnetization switching curves as a function of the current-density ($J$) through CrN layer under different in-plane positive magnetic fields. At 100\,Oe of magnetic field, no obvious switching was observed. Further increasing magnetic field to 200\,Oe and onward, the magnetization of top ferromagnetic layer switches at $J$ $\sim$ 2.6\,MA/cm$^2$. For negative magnetic fields (Fig.\ref{Fig2}(e)), a sign reversal of switching curve, indicates the magnetization switching is due to the SOT. Notably, the measured switching current density in our device is lower or comparable to the several HM/FM hetero-structure such as Ta/CoFeB/MgO \cite{qiu2014angular}, Ta/CoFeB/MgO/Ta \cite{liu2012spin}, Pt/Co/AlO$_x$ \cite{mihai2010current}, Pt/Co bilayer \cite{liu2012current}, Ta/CoFeB/TaO$_x$ \cite{yu2014switching}, and Hf/CoFeB/MgO \cite{akyol2015current}. Figure\,\ref{Fig2}(f) illustrates the plot of the switching current density ($J_c$) at which magnetization switches, against the in-plane magnetic fields. The $J_c$ lowers with increasing the magnetic field, which is a characteristics of SOT switching \cite{miron2011perpendicular,lo2014spin}.  Since the current channel of the device is along [1-10] direction, therefore charge current flowing in [1-10] direction produces the spin current flowing in [111] direction with spin-polarization along [11-2] direction. This spin-polarization exert a damping-like field on the magnetization of top ferromagnetic layer along [1-10] direction. When an external magnetic field is applied in [1-10] direction, the deterministic switching take place due to interfacial symmetry-breaking. It is important to note that our studied device contains Pt, which may facilitate magnetization switching within the ferromagnetic layer, a phenomenon known as self-induced SOT switching. A detailed discussion on self-induced SOT within the multilayer structure is provided in the Appendix of the manuscript.   
\vspace{-1mm}
\begin{figure}[htbp]
\centering
\includegraphics[width=0.5\textwidth]{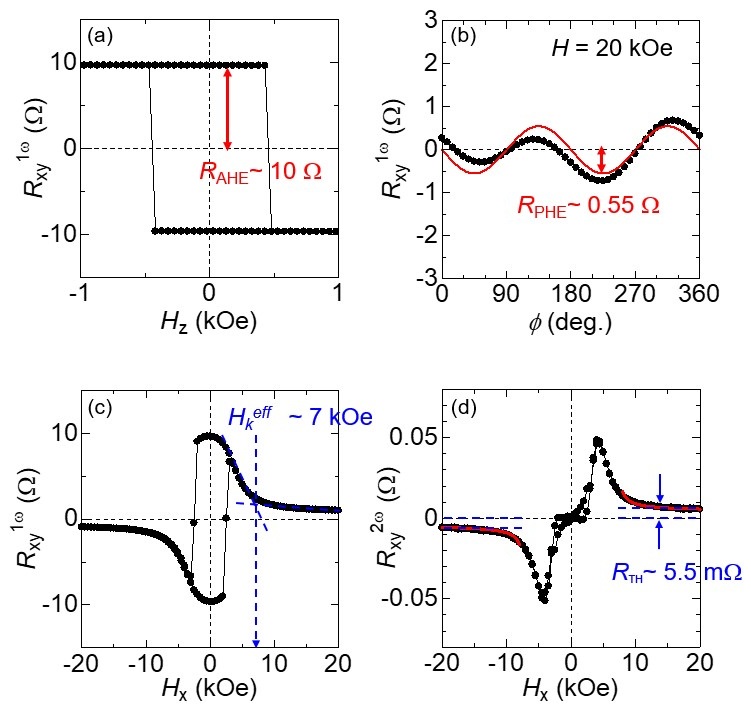}
\caption{(a) First-harmonic AHE loop (\textit{R}$_{xy}$$^{1\omega}$) for the out-of-plane magnetic field (\textit{H}$_{z}$). (b) Dependence of \textit{R}$_{xy}$$^{1\omega}$ on the in-plane angle of magnetic field ($\phi$) with respect to the current flow direction. (c) \textit{R}$_{xy}$$^{1\omega}$ for the in-plane magnetic field (\textit{H}$_{x}$). (d) Second-harmonic AHE loop (\textit{R}$_{xy}$$^{2\omega}$) for the \textit{H}$_{x}$. The red curves in Figs.\ref{Fig3}(b) and 3(d) correspond to the fitting results by Eq.\ref{eq:3} and Eq.\ref{eq:4}, respectively.}
\label{Fig3}
\end{figure}

In order to extract the damping-like SOT field and experimental SHC, we performed the second-harmonic Hall measurement \cite{yasuda2017current}, in which we applied AC current of 2.3 mA (3.3 MA/cm$^2$) to measure the first- (\textit{R}$_{xy}$$^{1\omega}$) and second-harmonic Hall resistances (\textit{R}$_{xy}$$^{2\omega}$) \cite{guo2021current}. Figure\ref{Fig3}(a) shows the \textit{R}$_{xy}$$^{1\omega}$ as a function of magnetic field in the out-of-plane direction (\textit{H}$_z$), suggesting the magnetic easy-axis points in the out-of-plane direction. The amplitude \textit{R}$_{AHE}$ was measured to be 10\,$\Omega$. Figure\,{\ref{Fig3}}(b) shows the \textit{R}$_{xy}$$^{1\omega}$ as a function of in-plane field angles ($\phi$) with respect to the AC current direction. The planar Hall resistance (\textit{R}$_{xy}$$^{1\omega}$)-$\phi$ can be expressed as \cite{itoh2019stack,lau2017spin}
\begin{equation}\label{eq:3}
    R_{xy}^{1\omega} = R_{PHE}sin2\phi.
\end{equation}
The \textit{R}$_{PHE}$ is found 0.55\,$\Omega$ by the fitting with Eq.\ref{eq:3} (red curve in Fig.\ref{Fig3}(b)), which is 20 times smaller than the \textit{R}$_{AHE}$ of 10\,$\Omega$ (Fig.\ref{Fig3}(a)). Figure \ref{Fig3}(c) shows the \textit{R}$_{xy}$$^{1\omega}$ as a function of in-plane field (\textit{H}$_{x}$). The $H_{k}^{eff}$ was measured to be 7\,kOe, which is consistent with that obtained in Fig.\ref{Fig1}(d). Figure\,\ref{Fig3}(d) shows the $H_x$-dependent $R_{xy}^{2\omega}$ up to field of 20\,kOe. To extract the damping like-torque field, the $R_{xy}$$^{2\omega}$ data was fitted using equation \cite{shirokura2021angle}
\begin{equation}\label{eq:4}
   R_{xy}^{2\omega} = \frac{R_{AHE}}{2}\frac{H_{DL}}{|H_x-H_{eff}|} + R_{PHE} \frac{H_{FL+Oe}}{|H_x|}+R_{TH},
\end{equation}
for the high field (shown in the red curve in  Figure\,\ref{Fig3}(d)). $R_{TH}$ and $H_{FL+Oe}$ are the resistance originating from the magneto-thermoelectric effect and the $H_{FL}$ that is contaminated by Oersted field, respectively. The $R_{TH}$ comprises of contribution in the $R_{xy}$ due to ordinary Nernst effect (ONE), anomalous Nernst effect (ANE) and spin Seebeck effect (SSE). Using $R_{{AHE}({PHE})}$ $\sim$ 10\,$\Omega$ (0.55 $\Omega$) and $H^{eff}_k$ $\sim$ 7\,kOe, we found $H_{DL}$= 135\,A/m. Note that $R_{TH}$ corresponds to the offset with respect to $R_{xy}^{2\omega}$ $\sim$ 0 at high field ($\sim$2\,T) [Fig.\ref{Fig3}(d)] is 5.5 m$\Omega$, suggesting that the magneto-thermoelectric effect is minor. Using $H_{DL}$, we calculated the effective spin Hall angle ($\theta^{eff}_{SH}$) using equation \cite{guo2021current}
\begin{equation}\label{eq:5}
    \theta^{eff}_{SH} = \frac{2e}{\hbar}\frac{\mu_0M_stH_{DL}}{J_{CrN}},
\end{equation}
where e, $\hbar$, $M_s$, $t$ and $J$ represent the electronic charge, reduced plank constant, saturation magnetization of top ferromagnetic layer, and current density through the CrN layer, respectively.  
From above equation using \( H_{DL} = 135 \, \text{A/m} \),  $t$ = 1.95\,nm , \( M_s = 8.9 \times 10^5 \, \text{A/m} \) and \( J_{CrN} = 0.5 \times 10^{10} \, \text{A/m}^2 \), we estimated \( \theta^{eff}_{SH} \) to be 0.18. The value of $\theta^{eff}_{SH}$ is used to calculate SHC using  $\frac{\hbar}{2e}\frac{\theta^{eff}_{SH}}{\rho}$, where $\rho$ is the resistivity of the CrN layer\cite{shi2021composition}. We found SHC ($\sigma^y_{zx}$) to be 83 ($\frac{\hbar}{e}$) S/cm.  Here $x$\,[1-10], $y$\,[11-2] and $z$\,[111] represent the direction of charge current, spin-polarization and spin-current, respectively. We calculated the $\theta_{SH}$ using the formula $\theta_{SH} = \frac{e}{\hbar} \frac{\sigma^y_{zx}}{\sigma_{xx}}$. Using $\sigma_{xx}$ = 917 S/cm from experiment, we calculated $\theta_{SH}$ $\sim$ 0.09, highlighting efficient charge-to-spin conversion in CrN. 
Since the spin-current from CrN switches the magnetization of the top ferromagnetic layer, it is essential and intriguing to investigate the origin of SHE in CrN using first-principles calculations.  
\begin{figure}[t]
    \centering
    \includegraphics[width=0.5\textwidth]{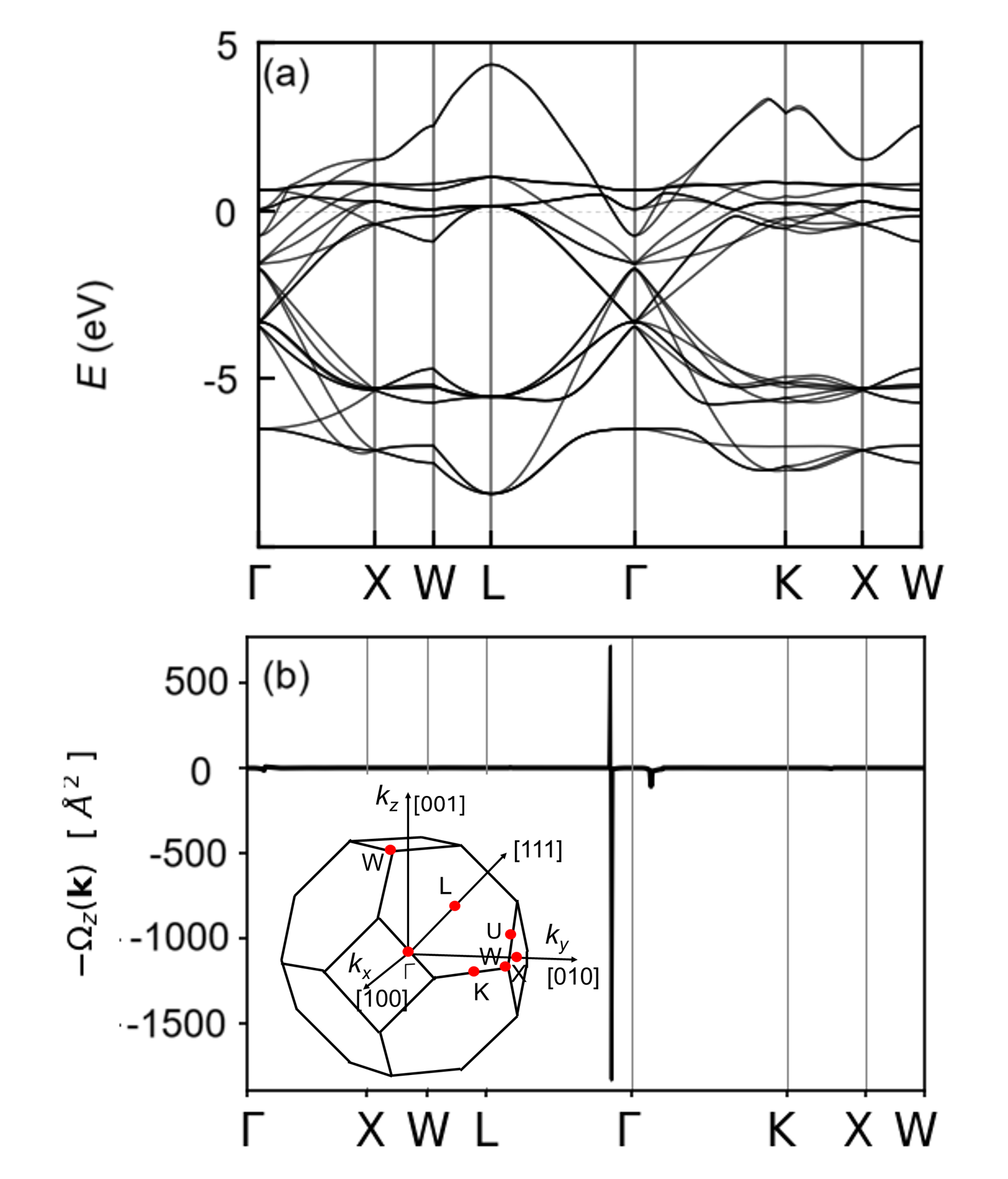}
    \caption{ (a) Band structure of CrN in presence of SOC. (b) Berry-curvature distribution along the same high-symmetry $k$-path. Inset shows a Brillouin zone of FCC lattice.}
    \label{Fig4}
\end{figure}

The first-principles calculation was performed keeping $z$-axis of the CrN unit cell in calculation along [111] direction and hence corresponding orthogonal $x$ and $y$ axis along [1-10] and [11-2] directions, respectively, to match with experimental situations (as CrN grow in (111) orientation on Al$_2$O$_3$ substrate). The Atomsk software was used for rotating the unit cell \cite{hirel2015atomsk}. Figure \ref{Fig4}(a) shows the band structure of CrN with SOC along the high-symmetry $k$-path in the  FCC Brillouin zone. A FCC Brillouin zone is shown in the inset of Fig.\ref{Fig4}(b). Around the $\Gamma$ high-symmetry $k$-point (near the Fermi energy), the bands shows degeneracy commonly refers to band splitting \cite{shukla2022band} due to SOC.  Since the transport properties mainly influence by the state close the Fermi energy, this band splitting could give the Berry curvature in the momentum space. For this, we computed $k$-resolved Berry curvature as shown in Fig.\ref{Fig4}(b) using Eq.\ref{eq:6}.  The Berry-curvature is negligible along the high symmetry $k$-path except around $\Gamma$ point, where it shows spike like behavior due to SOC induced band splitting. When the SOC-induced gap arises in the band structure, the denominator of Eq.\ref{eq:6} becomes small, leading to the emergence of Berry curvature in the material, which is inversely proportional to the magnitude of the SOC-induced gap. Since the device has a current channel along the mirror symmetry line (i.e., the $x$[1-10] direction), the charge current in the [1-10] direction generates a spin current flowing through the $z$[111] direction with spin polarization along the $y$[11-2] direction. Therefore our interest of calculation is $y$-component of SHC ($\sigma^y_{zx}$). Integration of Berry curvature in whole Brillouin zone using Eq.\ref{eq:7} gives $\sigma^y_{zx}$ $\sim$ 120 $\frac{\hbar}{e}$ S/cm at the Fermi energy. This SHC value is close to the obtained from second harmonic Hall measurement, which is responsible for magnetization switching in the device.  

 In discussion, we aimed to modify the electronic band structure and  Berry curvature to improve the SHE of Cr by incorporating N. As a result, we achieved the $\theta^{eff}_{SH}$ of 0.18 for the CrN-based SOT device, surpassing the $\theta^{eff}_{SH}$ of 0.088 reported for the Cr-based SOT device (Cr/CoFeB/MgO/Cr) \cite{chuang2019cr}. Our findings reveal that the increase in $\theta^{eff}_{SH}$ originates from the intrinsic enhancement of SHE, rather than extrinsic mechanism. The spin-current originating from CrN switches the magnetization of adjacent ferromagnet at a switching current density of 2.6 MA/cm$^2$, which is comparable to the existing HMs/FM systems. The agreement between experimental and theoretical SHC confirms the intrinsic Berry curvature origin of the spin current in CrN and validates the fundamental significance of our findings. In addition to the intrinsic mechanism for the impact from N, some reports suggest the extrinsic contribution. Xu \textit{et al.} studied the impact of N incorporation in Pt, and found the increased $\theta^{eff}_{SH}$ from 0.12 to 0.16 by the N ratio of 8\%. Spin-dependent scattering at the interface is considered to be one of the causes, where N incorporation improves interfacial spin transparency and reduces effective magnetic damping \cite {xu2021large}. Shashank \textit{et al.} suggested that increasing dose of N into Pt enhances the damping-like efficiency due to the extrinsic side jump mechanism \cite{shashank2023disentanglement}. As mentioned above, it is revealed that N incorporation into metals can provide a significant impact on their SHE owing to both intrinsic and extrinsic contributions. Our results open an alternative approach to enhance the SHE even when the value of pristine metals is not sufficient, as in the case of light $3d$ metals.
\section{CONCLUSION}
 We experimentally demonstrated CIMS in the Al$_2$O$_3$/CrN(5)/[Co(0.35)/Pt(0.3)]$_3$/MgO(3) multilayer device at a switching current density $\sim$ 2.6 MA/cm$^2$, in the presence of an in-plane external magnetic field. Theoretical calculations give the SHC ($\sigma^y_{zx}$) $\sim$ 120 $\frac{\hbar }{e}$ S/cm due to Berry curvature originating from SOC-induced band splitting, which is close to the experimental SHC. We found $\theta_{SH}$ to be 0.09, demonstrating efficient charge-to-spin conversion in CrN. Our work shows that light-element-based systems could be promising for energy-efficient and cost-effective SOT devices, even though they possess low SOC.
\section*{ACKNOWLEDGMENT}
This work was supported by KAKENHI Grants-in-Aid No. 23K22803 from the Japan Society for the Promotion of Science (JSPS). Part of this work was carried out under the Cooperative Research Project Program of the RIEC, Tohoku University.

These corresponding authors equally contributed to this work\\
\textsuperscript{*} shukla.gauravkumar@nims.go.jp\\
\textsuperscript{\dag}isogami.shinji@nims.go.jp

\par
\medskip
\appendix
\label{Appendix}
\renewcommand{\theequation}{A.\arabic{equation}}
\renewcommand{\thefigure}{A.\arabic{figure}}
\setcounter{equation}{0}
\setcounter{figure}{0}
\section*{{Appendix}}
The SHE arises due to SOC of the material and becomes more pronounced for HMs such as Pt, Ta, and W. Since top ferromagnetic layer [Co(0.35)/Pt(0.3)]$_3$ of the studied device consist of Pt, therefore the spin-current from the Pt may induced SOT and facilitates magnetization switching in the ferromagnetic layer, a phenomenon alternatively called self-induced SOT switching. 
\begin{figure}[b]
\centering
 \includegraphics[width=0.5\textwidth]{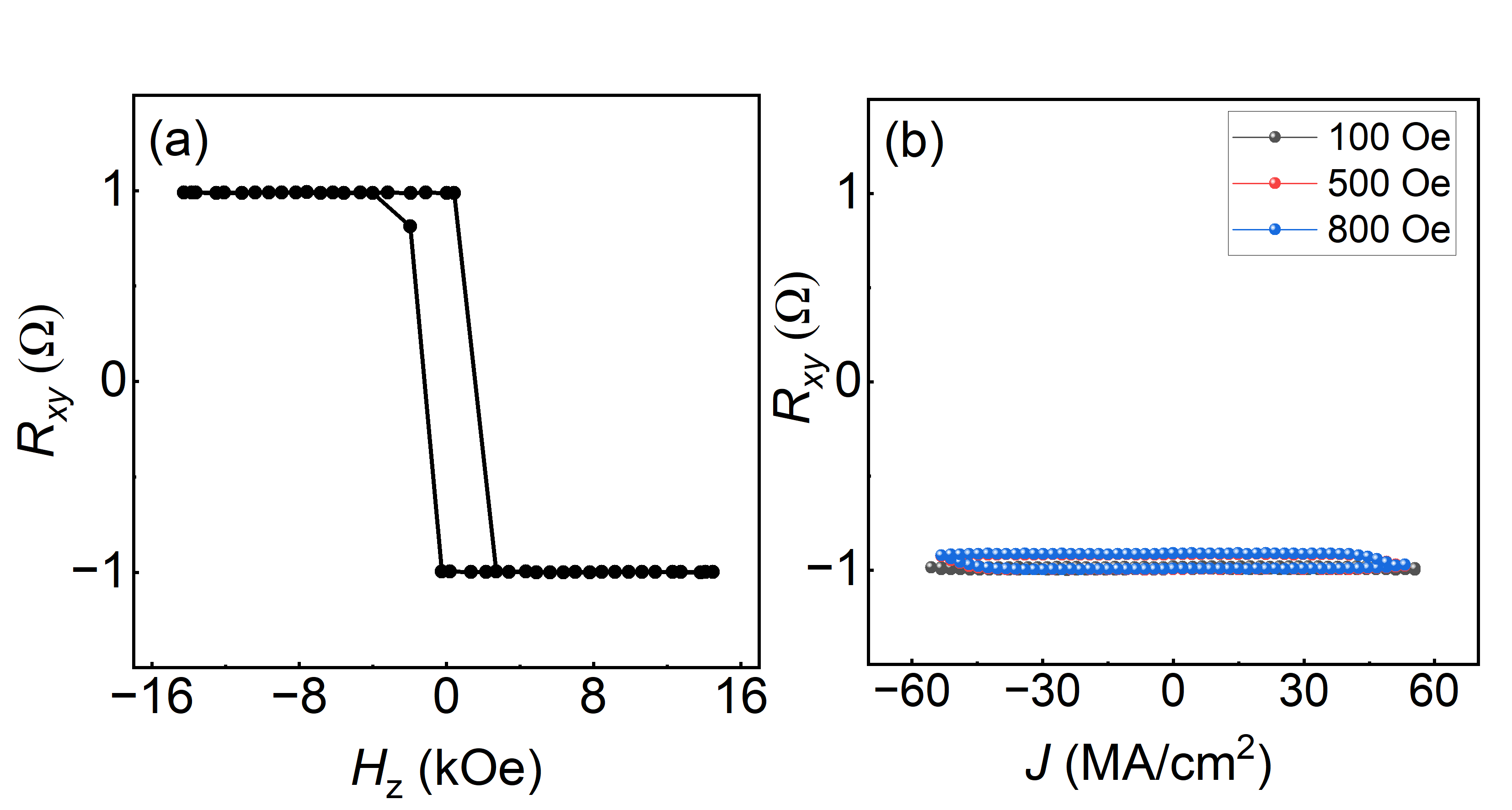}
\caption{ (a) Out-of-plane field-dependent anomalous Hall resistance $R_{xy}$. (b) Current-induced magnetization switching curve in the presence of different in-plane magnetic fields.}
\label{Fig:A1}
\end{figure}
However, the studied multilayer is vertically symmetric $i.e.$ there is no vertical composition gradient in top ferromagnetic layer, therefore the torque induced by top Pt layer on Co will be be canceled out by bottom Pt layer. Thus, the chance of the self-induced SOT is minimal in the device. Nevertheless, to further confirm whether self-induced SOT is present or not in the device, we synthesized [Co(0.35)/Pt(0.3)]$_6$ multilayer on $c$-plane oriented Al$_2$O$_3$ substrate without CrN layer. We fabricated Hall bar device and successfully achieved PMA in the device as shown in $R_{xy}$ versus out-of-plane magnetic field curve (Fig.\ref{Fig:A1}(a)). 
 Next we measured the CIMS in the presence of in-plane magnetic field of 100 Oe, 500 Oe and 800 Oe. The CIMS curve versus current density through the ferromagnetic layer is presented in Fig.\ref{Fig:A1}(b). From the switching curve we found that very small switching happens with in the ferromagnetic layer [Co(0.35)/Pt(0.3)]$_6$. This further confirms that vertically symmetric structure nullify the self-induced SOT and  the magnetization switching in our studied multilayer device Al$_2$O$_3$/CrN(5)/[Co(0.35)/Pt(0.3)]$_3$/MgO(3) is due to spin-current from CrN layer. 

\end{document}